\documentclass[reqno,a4paper,12pt,doublespace]{amsart}

\usepackage{amsfonts}

\usepackage{amssymb}

\usepackage{amsmath}

\usepackage{amsthm}

\newcommand{\Space}[2]{ \mathbb{#1}^{#2} }

\newcommand{\ub}[3][]{\left\{\!#1\left[#2,#3\right]\!#1\right\}}

\newtheorem{Thm}{Theorem}[section]

\newtheorem{Defn}{Definition}[section]

\newtheorem{Lemma}{Lemma}[section]

\newtheorem{Example}{Example}[section]

\newcommand{\Partial}[1]{ \frac{\partial}{\partial #1} }

\newcommand{\Fracdiffl}[2]{\frac{d #1}{d #2}}

\newcommand{\Heisn}{\Space{H}{n}}

\newcommand{\heisn}{\mathfrak{h}_{n}}

\newcommand{\uorb}{\mathcal{O}_{h}}

\newcommand{\hilbh}{\mathcal{H}_{h}}

\newcommand{\loneh}{L^1(\Heisn)}

\newcommand{\ltworn}{L^2(\Space{R}{n})}

\newcommand{\fock}{F^2(\uorb)}

\newcommand{\antid}{\mathcal{A}}

\newcommand{\zerodelxone}{\delta(s) \delta^{(1)}(x) \delta(y) }

\newcommand{\zerodelyone}{\delta(s) \delta(x) \delta^{(1)}(y) }

\newcommand{\vqp}{v_{(h,q,p)}}

\newcommand{\vqpd}{v_{(h,q',p')}}

\newcommand{\vqpdd}{v_{(h,q'',p'')}}

\newcommand{\qporb}{\mathcal{O}_{(q,p)}}

\newcommand{\proj}{\mathcal{P}}

\newcommand{\kct}{\mathcal{U}}

\newcommand{\lqp}{l_{(h,q,p)}}

\newcommand{\lqpm}{l_{(h,-q,-p)}}

\newcommand{\lqpp}{l_{(h,q',p')}}

\newcommand{\lqpd}{l_{(h,q',p')}}

\newcommand{\lqpo}{l_{(0,q,p)}}

\newcommand{\lqpom}{l_{(0,-q,-p)}}

\newcommand{\kctsp}{\kct^{*^+}}

\newcommand{\kctsm}{\kct^{*^-}}

\newcommand{\kctst}{\widetilde{\kct^*}}

\newcommand{\kctstcl}{\widetilde{\kct^{class}}}

\newcommand{\kctt}{\widetilde{\kct}}

\newcommand{\sokh}{\mathcal{L}_h}

\begin{document}

\date{19/05/04}

\title[Non-Linear Canonical Transformations]{Non-Linear Canonical Transformations in Classical and Quantum Mechanics}

\author{Alastair Brodlie}

\address{%
School of Mathematics\\
University of Leeds\\
Leeds LS2\,9JT\\
UK}

\email{abrodlie@amsta.leeds.ac.uk}

\maketitle

\begin{abstract}
$p$-Mechanics is a consistent physical theory which describes both classical and quantum mechanics simultaneously through the representation theory of the Heisenberg group. In this paper we describe how non-linear canonical transformations affect $p$-mechanical observables and states. Using this we show how canonical transformations change a quantum mechanical system. We seek an operator on the set of $p$-mechanical observables which corresponds to the classical canonical transformation. In order to do this we derive a set of integral equations which when solved will give us the coherent state expansion of this operator. The motivation for these integral equations comes from the work of Moshinsky and a variety of collaborators. We consider a number of examples and discuss the use of these equations for non-bijective transformations.
\end{abstract}

\vspace{0.5cm}

{\bf Keywords:} Canonical transformations, quantum mechanics, Heisenberg group, coherent states.

\vspace{0.5cm}

{\bf 2000 Mathematics Subject Classification:} Primary 70H15; Secondary 81S30, 81R30, 81R05.

\newpage

\section{Introduction}

Canonical transformations are at the centre of classical mechanics \cite{Arnold90,Goldstein80,Jose98}. A canonical transformation in classical mechanics is a map $A$ defined on phase space (throughout this paper we take phase space to be $\Space{R}{2n}$) which preserves the Poisson bracket. That is $A:\Space{R}{2n} \rightarrow \Space{R}{2n}$ such that for any two classical mechanical observables $f,g$
\begin{equation} \label{eq:canonpbcond}
\{ f \circ A , g \circ A \} = \{f,g\} \circ A.
\end{equation}
It is important to note that the map $A$ may well be non-bijective and non-linear. A condition which is equivalent to (\ref{eq:canonpbcond}) is that the map $A$ must also preserve the symplectic form on $\Space{R}{2n}$
\begin{equation}
\omega ( A(q,p), A(q',p')) = \omega ( (q,p),(q',p'))
\end{equation}
where $\omega$ is defined as $\omega((q,p),(q',p')) = qp'-q'p$. The most advanced applications of canonical transformations in classical mechanics are the Hamilton-Jacobi theory \cite[Chap. 10]{Goldstein80}, \cite[Chap. 9]{Arnold90} and action angle variables \cite[Sect. 6.2]{Jose98}, \cite[Chap. 9]{Arnold90}.

The passage of canonical transformations from classical mechanics to quantum mechanics has been a long journey which is still incomplete. The first person to give a clear formulation of quantum canonical transformations was Dirac, this is presented in his book \cite{Dirac47}. Mario Moshinsky along with a variety of collaborators has published a great deal of enlightening papers on the subject \cite{MelloMoshinsky75,MoshinskySeligman78,MoshinskySeligman79,GarciaMoshinsky80,DirlKasperkovitzMoshinsky88}. In these papers the aim is to find an operator, $U$, defined on a Hilbert space which corresponds to the canonical transformation. Moshinsky and his collaborators developed a system of differential equations which when solved gave the matrix elements --- with respect to the eigenfunctions of the position or momentum operator --- of $U$. More recently Arlen Anderson \cite{Anderson93,Anderson94} has published some results on modelling canonical transformations in quantum mechanics using non-unitary operators.

In this paper we use $p$-Mechanics to exhibit relations between classical and quantum canonical transformations. $p$-Mechanics \cite{KisilBrodlie02} describes both classical and quantum mechanics using the Heisenberg group (denoted $\Heisn$). The theory contains both observables and states which can both be realised as functions/distributions on $\Heisn$. $p$-Mechanical observables can be transformed into both quantum and classical observables using different representations of $\Heisn$.

We derive a system of integral equations using $p$-mechanics which when solved give the coherent state expansion of an operator on the set of $p$-mechanical observables corresponding to the canonical transformation. Under representations of $\Heisn$ this will give us the representation of canonical transformations in both classical and quantum mechanics. Our approach, unlike Moshinsky's,  does not need observables to be members of the algebra generated by the position and momentum operators.

In section \ref{sect:introtopmech} we give an outline of $p$-mechanics and extend it to fit the needs of this paper. In section \ref{sect:nonlineartrans} we derive systems of integral equations
for canonical transformations which when solved will give the corresponding operator on $p$-mechanical states in terms of coherent state expansions. For Hilbert space states this is presented in section \ref{sect:nonlineartransforhhstates} while for states realised as integration kernels these equations are derived in section
\ref{sect:nonlintransforstatekerns}. We consider applications of these equations to non-bijective transformations in section \ref{sect:nonbijcts}. Finally we summarise the paper and suggest some interesting extensions in section \ref{sect:summofctpaper}.

\section{$p$-Mechanics} \label{sect:introtopmech}

The theory of $p$-mechanics has been presented in a number of papers \cite{Brodlie02,Kisil02.1,Kisil02} --- a recent review article is \cite{KisilBrodlie02}. In this section we extend these concepts to fit the purposes of our paper. In particular we give a new definition of $p$-mechanical observables and show how the kernel states can be expanded out using coherent states.

At the heart of $p$-mechanics is the Heisenberg group \cite{Folland89, Taylor86}.
 The Heisenberg group (denoted $\Space{H}{n}$) is the set of all triples in $\Space{R}{} \times \Space{R}{n} \times \Space{R}{n}$ under the law of multiplication
\begin{equation}
(s,x,y)\cdot(s',x',y') = (s+s' + \frac{1}{2} (x \cdot y' -x' \cdot y),x+x',y+y').
\end{equation}
The non-commutative convolution of two functions $B_1,B_2 \in \loneh$ is
\begin{equation} \label{eq:convonhn}
(B_1 * B_2)(g) = \int_{\Heisn} B_1 (h) B_2 (h^{-1} g) dh = \int_{\Heisn} B_1 (g h^{-1}) B_2 (h) dh,
\end{equation}
where $dh$ is Haar measure on $\Heisn$ which is Lebesgue measure $ds \, dx \, dy$ on $\Space{R}{2n+1}$. In this paper the convolution algebra $\loneh$ is too restrictive so we extend convolution to spaces of distributions. Spaces of interest in this paper are:
\begin{itemize}
\item  $\mathcal{E}' (\Heisn)$ of distributions with compact support \cite[Thm. 24.2]{Treves67} on $\Heisn$;
 \item $\mathcal{S}' (\Heisn)$ of tempered distributions (also known as the Schwartz space) \cite[Defn. 25.2]{Treves67} on $\Heisn$;
\item $\mathcal{D}' (\Heisn)$ of all distributions \cite[Chap. 21]{Treves67} on $\Heisn$.
\end{itemize}
The convolution of two distributions is defined in a natural way \cite[Chap. 0]{Taylor86}. The spaces $\mathcal{E}'(\Heisn)$ and $\mathcal{D}'(\Heisn)$ are closed under convolution.

The Lie Algebra of $\Heisn$ is denoted by $\heisn$ and can be realised by the left invariant vector fields
\begin{displaymath}
\begin{array}{ccc}
S=\Partial{s}, & X_j = \Partial{x_j} - \frac{y_j}{2} \Partial{s}, & Y_j = \Partial{y_j} + \frac{x_j}{2} \Partial{s},
\end{array}
\end{displaymath}
with the Heisenberg commutator relations
\begin{equation} \nonumber
[ X_i , Y_j ] = \delta_{ij} S.
\end{equation}

The most common representation of the Heisenberg group is the Schr\"odinger representation \cite[Sect.1.3]{Folland89}, \cite[Eq. 2.23]{Taylor86} on $\ltworn$
\begin{equation} \label{eq:schrorep}
\left( \rho_h^S (s,x,y) \psi \right) (\xi) = e^{-2\pi ihs - 2\pi ix\xi -\pi ihxy} \psi(\xi+hy).
\end{equation}
 Throughout this paper we do not use (\ref{eq:schrorep}), instead we show how a represenation unitarily equivalent to this can at times be advantageous. We now introduce this representation and the space on which it is defined.
\begin{Defn}
We define the space $\fock$ as
\begin{equation} \label{eq:defoffock}
\fock = \{ f_h (q,p) \in L^2 (\Space{R}{2n})  : D^j_h f_h =0 ,\textrm{ for } 1 \leq j \leq n \} ,
\end{equation}
where the operator $D_h^j$ on $L^2 (\Space{R}{2n})$ is defined as
$\frac{h}{2} \left( \Partial{p_j} + i \Partial{q_j} \right) + 2 \pi ( p_j +i q_j)$.
\end{Defn}
The inner product on $\fock$ is given by
\begin{equation} \label{eq:fockip}
\langle v_1, v_2 \rangle_{\fock} = \left( \frac{4}{ h } \right)^{n} \int_{\Space{R}{2n}} v_1 (q,p) \overline{v_2 (q,p)} \, dq \, dp
\end{equation}
$\fock$ is a Hilbert space with this inner product \cite{Kisil99}[Sect. 4.1]. The motivation for using this space in $p$-mechanics originates from Kirillov's method of orbits  \cite{Kirillov76,Kirillov99} --- this relation is discussed in \cite{KisilBrodlie02,Kisil02.1}. $\fock$ is similar to the Fock-Segal-Bargmann (\cite{Bargmann61}, \cite[Sect. 1.6]{Folland89} , \cite[Chap. 1]{Taylor86})  space of analytic functions on $\Space{C}{n}$ which are square integrable with respect to the measure $e^{-2|z|^2 /h} \, dz$. It is shown in \cite[Prop. 2.6]{Kisil02.1} that $f_h (q,p)$ is in $\fock$ if and only if $f_h(z)e^{|z|^2/h}$ is in the Fock-Segal-Bargmann space with $z=p+iq$. The integral kernel
\begin{equation} \nonumber
K_I (q,p,x) = e^{2\pi i qx - \pi i pq} e^{-\pi (x-p)^2}
\end{equation}
provides an isometry $W:\ltworn \rightarrow \fock$ by
\begin{equation} \label{eq:isomtofock}
\psi (x) \mapsto f(q,p) = \int_{\Space{R}{n}} \psi (x) K_I (q,p,x)dx.
\end{equation}
This is proved in \cite[Sect. 4.2]{Kisil99}. It is also shown in \cite[Sect. 4.2]{Kisil99} that $\fock$ is a reproducing kernel Hilbert space with reproducing kernel
\begin{eqnarray} \nonumber
\lefteqn{K_R (q,p,q',p')} \\ \nonumber
&& \hspace{0.2cm} = \exp \left( -\frac{2\pi}{h} \left( q^2 +p^2 +q'^2 + p'^2 - 2qq' -2pp' -2iq'p +2iqp' \right) \right).
\end{eqnarray}

The representation $\rho_h$ \cite{Kisil02.1,KisilBrodlie02} of $\Heisn$ on $F^2 (\uorb)$ is defined by
\begin{equation} \label{eq:infdimrep}
\rho_h (s,x,y): f_h (q,p) \mapsto e^{-2 \pi i (hs + qx +py)} f_h(q-\frac{h}{2} y, p+\frac{h}{2} x),
\end{equation}
which is unitary with respect to the inner product defined in (\ref{eq:fockip}). This representation is intertwined with the Schr\"odinger representation by the unitary map (\ref{eq:isomtofock}) \cite{Kisil99} and so is unitarily equivalent to the Schr\"odinger representation.

The crucial theorem which motivates the whole of $p$-mechanics is
\begin{Thm} \nonumber
(The Stone-von Neumann Theorem) All unitary irreducible representations of the Heisenberg group, $\Space{H}{n}$, up to unitary equivalence, are either:

(i) of the form $\rho_h$ on $\fock$ from equation (\ref{eq:infdimrep}), or

(ii) for $(q,p) \in \Space{R}{2n}$ the commutative  one-dimensional representations on $\Space{C}{} = L^2 (\qporb )$
\begin{equation} \label{eq:onedimrep}
\rho_{(q,p)} (s,x,y)u = e^{- 2 \pi i(q.x + p.y)} u.
\end{equation}
\end{Thm}
\begin{proof}
In \cite{Folland89} or \cite{Taylor86} it is shown that this holds for the Schr\"odinger representation. Our result follows since $\rho_h$ is intertwinned with the Schr\"odinger representation by the isometry $W$ given in (\ref{eq:isomtofock}).
\end{proof}
We can extend both $\rho_h$ and $\rho_{(q,p)}$ to the representation of an infinitely differentiable compactly supported function, $B\in C^{\infty}_0 (\Heisn)$, on $\Heisn$ by
\begin{equation} \nonumber
\rho (B) = \int_{\Heisn} B(g) \rho(g) dg.
\end{equation}
The representation of distributions is done in the natural way \cite[Chap 0, Eq 3.4]{Taylor86}.

 The basic idea of $p$-mechanics is to choose particular functions or distributions on $\Heisn$ which under the infinite dimensional representation will give quantum mechanical observables, while under the one dimensional representation will give classical mechanical observables. In doing this it is shown that both mechanics are derived from the same source. $p$-Mechanical observables can be realised as operators (some of which are unbounded) on a subset of $L^2 (\Heisn)$ generated by convolutions of the chosen functions or distributions. To define $p$-mechanical observables properly we need to introduce a map from the set of classical observables to the set of $p$-mechanical observables. In \cite{KisilBrodlie02,Kisil02.1} a map of $p$-mechanisation, $\proj$, from the set of classical observables to the set of $p$-mechanical observables is defined as
\begin{equation} \label{eq:pmechmap}
(\proj f) (s,x,y) = \delta (s) \breve{f} (x,y)
\end{equation}
where $f$ is any classical observable and $\breve{f}$ is the inverse Fourier transform of $f$ (that is, $\breve{f} (x,y) = \int_{\Space{R}{2n}} f(q,p) e^{2\pi i (qx+py)} \, dq \, dp$).
\begin{Defn}[$p$-Mechanical Observables]
The set of $p$-mechanical observables is the image of the set of classical observables under the map $\proj$ from equation (\ref{eq:pmechmap}).
\end{Defn}
Clearly this definition depends on how the set of classical observables is defined. Any physically reasonable classical mechanical observable can be realised as an element of $\mathcal{S}' (\Space{R}{2n})$. Since the Fourier transfrom maps $\mathcal{S}'(\Space{R}{2n})$ into itself, $\mathcal{S}'(\Heisn)$ is a natural choice for the set of $p$-mechanical observables. It includes the image of all classical observables which are polynomials or exponentials of the variables $q$ and $p$.

If we take the $\rho_h$ representation (\ref{eq:infdimrep}) of many of the distributions described above we would get unbounded operators. For example the distribution $\zerodelxone$ under the $\rho_h$ representation will generate the unbounded operator $\frac{h}{2} \Partial{p} - 2\pi i q I$. This operator is clearly not defined on the whole of $\fock$. This technical problem can be solved by the usual method of rigged Hilbert spaces (also known as Gelfand triples) \cite{Bohm02,Roberts66,Ruelle66} which uses the theory of distributions. Another approach to dealing with unbounded operators is given by using the G\aa rding space as explained in \cite[Chap. 0]{Taylor86}.

The dynamics of a $p$-mechanical system is described in \cite{KisilBrodlie02,Kisil02,Kisil02.1} using the universal brackets.
The universal brackets (also known as $p$-mechanical brackets) are
\begin{equation} \label{eq:pmechbrackets}
\ub{B_1}{B_2}{} = \antid ( B_1 * B_2 - B_2 * B_1 )
\end{equation}
where $\antid$ is the right inverse to the vector field $S=\Partial{s}$. It is shown in \cite[Prop. 3.5]{Kisil02} that under the one and infinite dimensional representations the universal brackets become the Poisson brackets and the quantum commutator respectively. Hence for a system with Hamiltonian $B_H$ (the $p$-mechanisation of the classical Hamiltonian $H$) solving the $p$-dynamic equation
\begin{equation} \label{eq:pdyneqn}
\Fracdiffl{B}{t} = \ub{B}{B_H}{}
\end{equation}
will give the quantum and classical dynamics under the infinite and one dimensional representations respectively.

In \cite{Brodlie02,KisilBrodlie02}, states in $p$-mechanics were introduced. They were defined as functionals on the set of $p$-mechanical observables and came in two forms --- elements of a Hilbert space and integration kernels.
\begin{Defn}
The Hilbert space $\hilbh$, $h \in \Space{R}{} \setminus \{ 0 \}$, is defined as the set of functions on $\Heisn$
\begin{equation} \label{hh}
\hilbh = \left\{ e^{-2 \pi ihs} f (x,y) : E^j_h f = 0 \hspace{0.4cm} 1 \leq j \leq n \hspace{0.4cm} \textrm{and} \hspace{0.4cm} f \in L^2 (\Space{R}{2n}) \right\}
\end{equation}
where the operator $E^j_h = \pi h (y-ix) +i\Partial{x}-\Partial{y}$
(this is the Fourier transform of $D_h^j$ from (\ref{eq:defoffock})).
\end{Defn}
The inner product on $\hilbh$ is defined as
\begin{equation}  \label{hhip}
\langle v_1 , v_2 \rangle_{\hilbh} = \left( \frac{4}{h} \right)^{n} \int_{\Space{R}{2n}} v_1 (s,x,y) \overline{v}_2 (s,x,y) \, dx \, dy.
\end{equation}
The set of $p$-mechanical observables acts on $\hilbh$ by  convolution. For many observables this will give rise to unbounded operators which are not defined on the whole of $\hilbh$. This problem is solved as before by the use of rigged Hilbert spaces. It is shown in \cite[Eq. 3.4]{Brodlie02} that any element $v \in \hilbh$ is of the form $v(s,x,y) = e^{-2\pi ihs} \hat{f} (x,y)$ for some $f \in \fock$ ($\hat{f}$ denotes the Fourier transform of $f$).

The state corresponding to $v \in \hilbh$ can be realised by an integration kernel
\begin{equation} \label{eq:relationbetweenkernelandvector}
l(s,x,y) = \left(\frac{4}{h}\right)^n \int_{\Space{R}{2n}}  v((s,x,y)^{-1} (s',x',y')) \overline{v((s',x',y'))} \, dx' \, dy' .
\end{equation}
For any $p$-mechanical observable $B$ the following  relation is proved in \cite[Thm. 3.1, Thm. 3.2]{Brodlie02}
\begin{equation} \label{eq:expvalueofobsinstate}
\langle \rho_h (B) f,f \rangle = \langle B*v,v \rangle = \int_{\Heisn}B(g) l(g) \, dg
\end{equation}
where $f$ is the element of $\fock$ such that $v(s,x,y) = e^{-2\pi ihs} \hat{f}(x,y)$  and $l$ is the kernel corresponding to $v$ through relation (\ref{eq:relationbetweenkernelandvector}). (\ref{eq:expvalueofobsinstate}) gives the expectation value of the observable $B$ in the state corresponding to $f$,$v$ and $l$.

In \cite{Brodlie02} an overcomplete system of coherent states in $\hilbh$ are derived using representations of the Heisenberg group
\begin{eqnarray} \label{eq:hhcoherentstates}
\lefteqn{v_{(h,q,p)}(s,x,y) } \\ \nonumber
&=& \left( \frac{h}{2} \right)^n \exp \left( -2\pi i h s + \pi i( xq+yp) - \frac{\pi h}{2} \left( \left( x+\frac{p}{h} \right)^2 + \left( y-\frac{q}{h} \right)^2 \right) \right).
\end{eqnarray}
The corresponding kernel coherent states are
\begin{equation} \label{eq:lhcoherentstates}
\lqp =  \exp \left( -2\pi i (q x + p y) + 2\pi ihs -\frac{\pi h}{2} \left( x^2 + y^2 \right) \right).
\end{equation}
It is shown in \cite{Brodlie02} that if we choose $B= \proj (f)$ then
\begin{equation} \nonumber
\lim_{h \rightarrow 0} \langle B* \vqp , \vqp \rangle = \int_{\Heisn} B\, \lqp \, dg = f(q,p).
\end{equation}

By the usual theory of coherent states (that is, wavelets) \cite{AliAntGaz00}, in a Hilbert space any element $v \in \hilbh$ can be written as
\begin{equation} \label{eq:expofhhcs}
v = \int_{\Space{R}{2n}} \langle v, \vqp \rangle \vqp \, dq \, dp.
\end{equation}
If we define an inner product on the set of kernels as
\begin{equation} \label{eq:iponkers}
\langle l_1 , l_2 \rangle = \int_{\Space{R}{2n}} l_1 (s,x,y) \overline{l_2 (s,x,y)} \, dx \, dy.
\end{equation}
This inner product will be well defined for any two kernel coherent states from (\ref{eq:lhcoherentstates}) since the integral on the right hand side of (\ref{eq:iponkers}) will be finite. Then we can define our space of kernels, $\sokh$, as the completion of the set of linear combinations of the coherent states (\ref{eq:lhcoherentstates}). Clearly $\sokh$ is a Hilbert space so we can expand any kernel in $\sokh$ by the formula
\begin{equation} \nonumber
l(s,x,y) = \int_{\Space{R}{2n}} \int_{\Space{R}{2n}} l(s,x',y') \overline{l_{(h,q,p)} (s,x',y')} \, dx' \, dy' l_{(h,q,p)} (s,x,y) dq dp.
\end{equation}

We now show that the $\hilbh$ coherent states are eigenfunctions of the creation and annihilation operators. The creation and annihilation distributions are defined as
\begin{eqnarray} \label{eq:defaplus}
a^+ &=& \frac{1}{2\pi i} \left( \zerodelxone - i  \zerodelyone \right), \\ \label{eq:defaminus}
a^- &=& \frac{1}{2 \pi i} \left( \zerodelxone + i  \zerodelyone \right) .
\end{eqnarray}
The creation and annihilation operators are convolution by the above distributions. It should be noted that $a^+$ and $a^-$ are the $p$-mechanisation of the classical observables $q-ip$ and $q+ip$ respectively. By a direct calculation it can be shown that
\begin{equation} \nonumber
a^- * v_{(h,q,p)} = (q+ip) v_{(h,q,p)},
\end{equation}
so $v_{(h,q,p)}$ is an eigenfunction for $a^-$ with eigenvalue $(q+ip)$. By another direct calculation using (\ref{eq:waveletwithx}) we have
\begin{eqnarray} \label{eq:ipoftwocs}
\lefteqn{\langle v_{(h,q,p)} , v_{(h,q',p')} \rangle_{\hilbh}} \\ \nonumber
&& = \exp \left( -\frac{\pi}{2h} \left( (p-p')^2 + (q-q')^2 + 2i(qp' - q'p) \right) \right).
\end{eqnarray}
Finally by another direct calculation we have that $a^-$ and $a^+$ are adjoints of each other.

A well known equation which will be used throughout this paper is \begin{equation}
\label{eq:waveletwithx}
\int_{\Space{R}{}} \exp ( -a x^2 + 2b x ) \, dx = \left( \frac{\pi}{a} \right)^{\frac{1}{2}} \exp \left( \frac{b^2}{a} \right).
\end{equation}
where $a>0$. A similar equation \cite[p337]{GradshteynRyzhik80} which we repeatedly use is
\begin{equation} \label{eq:waveletwithxforalln}
\int_{\Space{R}{}} x^n \exp(-ax^2+2bx) \, dx = \frac{1}{2^{n-1} a} \left( \frac{\pi}{a}\right)^{\frac{1}{2}} \frac{d^{n-1}}{db^{n-1}} \left( b \exp \left( \frac{b^2}{a} \right) \right)
\end{equation}
providing $a>0$ and $n$ is an integer greater than or equal to $1$.  This equation for the particular value of $n=1$ is well known
\begin{equation} \label{eq:waveletwith}
\int_{\Space{R}{}} x \exp ( -a x^2 + 2b x ) \, dx = \left( \frac{\pi}{a} \right)^{\frac{1}{2}} \left( \frac{b}{a} \right) \exp \left( \frac{b^2}{a} \right).
\end{equation}

\section{Non-Linear Canonical Transformations} \label{sect:nonlineartrans}

In \cite{Brodlie02,KisilBrodlie02} the $p$-dynamic equation (\ref{eq:pdyneqn}) for the forced and harmonic oscillators are solved in $p$-mechanics. In doing so it was made evident that the quantum and classical pictures of the problems were generated from the same source. To solve the $p$-dynamic equation for more complicated problems, such as the Kepler problem, technical problems are encountered. In classical mechanics when these problems arise the solution often lies in finding a canonical transformation to a set of coordinates in which Hamilton's equations have a more manageable form. For example the transformation to action-angle variables completely solves the Kepler problem \cite[Sect. 10.8]{Goldstein80}. By studying canonical transformations in $p$-mechanics we have a tool which will transform the $p$-dynamic equation (\ref{eq:pdyneqn}) into a more desirable form.

In studying $p$-mechanical canonical transformations we show how canonical transformations can be represented in the mathematical framework of both quantum and classical mechanics. It is stated in \cite{Anderson94} that canonical transformations have three important roles in both quantum and classical mechanics:
\begin{itemize}
\item time evolution;
\item physical equivalence of two theories;
\item solving a system.
\end{itemize}
Taking the one and infinite dimensional representation of the $p$-mechanical system will show how these properties are exhibited in classical and quantum mechanics respectively.

There are further benefits of considering canonical transformations in $p$-mechanics. Canonical transformations can represent the symmetries of a classical mechanical system. In looking at the image of canonical transformations in quantum mechanics we can see how these symmetries are represented in quantum mechanics. In \cite{Anderson93} Anderson shows how quantum integrability can be defined in terms of canonical transformations. In \cite{Anderson94.1} it is shown that quantum canonical transformations can also help in the study of partial differential equations.

In this section we consider non-linear canonical transformations. The role of linear canonical transformations in $p$-mechanics is straightforward and is described in \cite{KisilBrodlie02,Kisil02.1,Kisil02.2}. Unfortunately some of the most fundamental canonical transformations are non-linear --- for example the passage to action angle variables for the harmonic and the repulsive oscillator.

For non-linear transformations we follow an approach which is an enhancement of a method pioneered by Mario Moshinsky and a variety of collaborators \cite{MelloMoshinsky75,MoshinskySeligman78,MoshinskySeligman79,GarciaMoshinsky80,DirlKasperkovitzMoshinsky88}. In this paper we are looking at general $p$-mechanical observables as opposed to just quantum mechanical observables. We also make use of the $p$-mechanical coherent states (\ref{eq:hhcoherentstates}), (\ref{eq:lhcoherentstates}).

\subsection{Equations for non-linear transformations involving $\hilbh$ states} \label{sect:nonlineartransforhhstates}

This method starts with the observation that a canonical transformation in classical mechanics described by $2n$ independent relations
\begin{eqnarray} \label{eq:oldinitialconrelq}
q_i \rightarrow Q_i(q,p) \\ \label{eq:oldinitialconrelp}
p_i \rightarrow P_i(q,p)
\end{eqnarray}
$i=1, \hdots , n$ where $\{ Q_i , P_j \}_{q,p}= \delta_{ij}$ can be realised by $2n$ functional relations
\begin{eqnarray} \label{eq:newinitialconrelq}
f_i (q,p) &=& F_i (Q,P) \\ \label{eq:newinitialconrelp}
g_i (q,p) &=& G_i (Q,P)
\end{eqnarray}
for $i=1, \hdots ,n$ where $\{ f_i,g_i\}_{q,p} = \{ F_i,G_i \}_{Q,P}$.
The advantage of this approach is that the $p$-mechanisation (\ref{eq:pmechmap}) of the functions in (\ref{eq:newinitialconrelq}, \ref{eq:newinitialconrelp}) may be easier to derive than the functions on the right hand side of equations (\ref{eq:oldinitialconrelq}, \ref{eq:oldinitialconrelp}). We assume throughout the paper that the above functions of $q$ and $p$ are $C^{\infty}$ with isolated singularities and when integrated next to an element of $\mathcal{S}(\Space{R}{2n})$ will be finite. This means they can always be realised as elements of $\mathcal{S}'(\Space{R}{2n})$. The isolated singularity condition means the equality in system (\ref{eq:newinitialconrelq}, \ref{eq:newinitialconrelp}) holds everywhere except at a finite number of isolated points.

We now derive an equation which will give us a clear form of an operator $U$ on $\hilbh$ corresponding to a canonical transformation. This equation will supply us with the matrix elements of the operator $U$ with respect to the overcomplete set of coherent states, that is  it will give us $\langle U v_{(h,q,p)} , v_{(h,q',p')} \rangle$ for all $q,p,q',p' \in \Space{R}{n}$.

In Dirac's original treatment of quantum canonical transformations \cite[Chap. 27]{Dirac47} he proposed that the canonical transformation from equations (\ref{eq:oldinitialconrelq}) and (\ref{eq:oldinitialconrelp}) should be represented in quantum mechanics by a unitary operator $U$ on a Hilbert space such that
\begin{equation} \nonumber
\tilde{Q_i}= U\tilde{q_i} U^{-1} \hspace{1cm} \textrm{and} \hspace{1cm} \tilde{P_i} = U \tilde{p_i} U^{-1} ,
\end{equation}
$i=1, \cdots ,n$. Here $\tilde{Q_i},\tilde{P_i},\tilde{q_i}, \tilde{p_i}$ are the quantum mechanical observables corresponding to the classical mechanical observables $Q_i,P_i,q_i,p_i$ respectively.

In \cite{MelloMoshinsky75} Mello and Moshinsky suggested that in some circumstances it is easier to define the operator $U$ by the equations
\begin{equation} \nonumber
\tilde{F} U = U\tilde{f}  \hspace{1cm} \textrm{and} \hspace{1cm} \tilde{G} U = U \tilde{g}
\end{equation}
where $\tilde{F}, \tilde{G}, \tilde{f}, \tilde{g}$ are the quantum mechanical observables (that is operators on a Hilbert space) corresponding to the classical observables $F,G,f,g$ from equations (\ref{eq:newinitialconrelq}) and (\ref{eq:newinitialconrelp}).

We proceed to transfer this approach into $p$-mechnaics. We want to understand the operator $U$ which is defined by the equations
\begin{eqnarray} \label{eq:beginofourmmeqn}
\proj ( f_i (q,p) ) * U v &=& U \proj (F_i(Q,P)) * v , \\ \label{eq:beginofourmmeqntwo}
 \proj ( g_i (q,p) ) * U v &=& U \proj (G_i(Q,P)) *v ,
\end{eqnarray}
where $\proj$ is the map of $p$-mechanisation (\ref{eq:pmechmap}) and $v$ is any element of $\hilbh$.

We will now divert from deriving the general equation by giving an example to illuminate these ideas (the example we give is a linear transformation but it must be stressed that this work holds for non-linear transformations too).
\begin{Example} \label{ex:flipexample}
Consider the linear canonical transformation
\begin{equation} \nonumber
q \rightarrow -P \hspace{2cm} p \rightarrow Q.
\end{equation}
This can be realised by the two equations
\begin{eqnarray} \label{eq:goodformofflipctone}
q+ip &=& -P+iQ \\ \label{eq:goodformofflipcttwo}
q-ip &=& -P-iQ
\end{eqnarray}
the $p$-mechanisation of which are
\begin{eqnarray} \label{eq:flipincreaandanihilone}
a^- &=& i A^-  \\ \label{eq:flipincreaandanihiltwo}
a^+ &=& -i A^+
\end{eqnarray}
where $a^-$ and $a^+$ are defined in equations (\ref{eq:defaplus}) and (\ref{eq:defaminus}).
\end{Example}

We now continue to derive the equation which will help us understand the operator $U$. For the rest of this section we just write the equations out using $f_i$ and $F_i$, but all these will still hold if they are replaced by $g_i$ and $G_i$. We begin by taking the matrix elements of equation (\ref{eq:beginofourmmeqn}) with respect to the coherent states defined in equation (\ref{eq:hhcoherentstates}); we get
\begin{equation} \label{eq:matrixeltofcteqn}
\langle \proj ( f_i) * U \vqp , \vqpd \rangle = \langle U \proj (F_i) * \vqp, \vqpd \rangle.
\end{equation}
We can expand $U \vqp$ using our system of coherent states
\begin{equation} \nonumber
U \vqp = \int_{\Space{R}{2n}} \langle U \vqp , \vqpdd \rangle \vqpdd \, dq'' \, dp''.
\end{equation}
The left hand side of equation (\ref{eq:matrixeltofcteqn}) now becomes
\begin{equation} \nonumber
\int_{\Space{R}{2n}} \langle U \vqp , \vqpdd \rangle \langle \proj (f_i) * \vqpdd, \vqpd \rangle \, dq'' \, dp''.
\end{equation}
Similarly we expand out $\proj (F_i) * \vqp$ out as
\begin{equation} \nonumber
\proj (F_i) * \vqp = \int_{\Space{R}{2n}} \langle  \proj (F_i) * \vqp, \vqpdd \rangle \vqpdd \, dq'' \,  dp''.
\end{equation}
so the right hand side of (\ref{eq:matrixeltofcteqn}) becomes
\begin{equation} \nonumber
\int_{\Space{R}{2n}} \langle U \vqpdd, \vqpd \rangle \langle \proj (F_i) *\vqp, \vqpdd \rangle \, dq'' \, dp''.
\end{equation}

Hence if we set $m(a,b,c,d) = \langle U v_{(h,a,b)} , v_{(h,c,d)} \rangle$ equation (\ref{eq:matrixeltofcteqn}) becomes
\begin{eqnarray} \label{eq:thefinaleqnforkernsinctsof}
\lefteqn{\int_{\Space{R}{2n}} m(q,p,q'',p'') \langle \proj (f_i) * \vqpdd, \vqpd \rangle \, dq'' \, dp''} \\ \nonumber
&& = \int_{\Space{R}{2n}} m(q'',p'',q',p') \langle \proj (F_i) *\vqp, \vqpdd \rangle \, dq'' \, dp''.
\end{eqnarray}
Note that to get the full system of equations we need a further $n$ equations which we get by replacing $f_i$ and $F_i$ with $g_i$ and $G_i$.
If we can solve this integral equation for $m$ then we can understand the effect of $U$ on any element $v$ of $\hilbh$ through coherent state expansions. Even if we can not solve the integral equation (\ref{eq:thefinaleqnforkernsinctsof}) we can still gain some useful insights into the nature of the canonical transformation in question.

By (\ref{eq:expofhhcs}) the unitarity of $U$ is equivalent to the following two equations holding
\begin{eqnarray} \nonumber
\lefteqn{\int_{\Space{R}{2n}} \langle v_{(h,q',p')} , U  v_{(h,q''',p''')} \rangle \langle U^T v_{(h,q'',p'')} , v_{(h,q''',p''')} \rangle \, dq''' \, dp'''} \\ \nonumber
&& \hspace{6cm} = \langle v_{(h,q',p')} , v_{(h,q'',p'')} \rangle \\ \nonumber
\lefteqn{\int_{\Space{R}{2n}} \langle v_{(h,q',p')} , U^T  v_{(h,q''',p''')} \rangle \langle U v_{(h,q'',p'')} , v_{(h,q''',p''')} \rangle \, dq''' \, dp'''} \\ \nonumber
&& \hspace{6cm}  = \langle v_{(h,q',p')} , v_{(h,q'',p'')} \rangle,
\end{eqnarray}
where $U^T$ stands for Hermitian conjugate.
Since for many functions $f$, $\langle \proj (f) *\vqp,\vqpd \rangle$ is a manageable function of $q,p,q',p'$ equation (\ref{eq:thefinaleqnforkernsinctsof}) will take a simple form for a variety of examples. For example consider the distributions involved in equations (\ref{eq:flipincreaandanihilone}) and (\ref{eq:flipincreaandanihiltwo}). Since $\vqp$ is an eigenfunction of the annihilation operator $a^-$ with eigenvalue $(q+ip)$ we have
\begin{equation} \nonumber
\langle a^- * \vqp, \vqpd \rangle = (q+ip) \langle \vqp, \vqpd \rangle.
\end{equation}
and hence
\begin{eqnarray} \label{eq:actiofcreatonctsone}
\lefteqn{\langle \proj (q+ip) * \vqp,\vqpd \rangle} \\ \nonumber
&=& (q+ip) \langle \vqp, \vqpd \rangle \\ \nonumber
&=& (q+ip) \exp \left( -\frac{\pi}{2h} \left( (p-p')^2 + (q-q')^2 + 2i(qp' - q'p) \right) \right)
\end{eqnarray}
here we have used (\ref{eq:ipoftwocs}). Furthermore since $a^-$ is the adjoint of $a^+$ we have
\begin{eqnarray} \nonumber
\langle a^+ * \vqp , \vqpd \rangle &=& \langle \vqp, a^- * \vqpd \rangle \\ \nonumber
&=& (q'-ip') \langle \vqp, \vqpd \rangle,
\end{eqnarray}
hence
\begin{eqnarray} \label{eq:actiofanihilonctsone}
\lefteqn{ \langle \proj (q-ip) * \vqp , \vqpd \rangle} \\ \nonumber
&=& (q'-ip') \langle \vqp, \vqpd \rangle \\ \nonumber
&=& (q'-ip') \exp \left( -\frac{\pi}{2h} \left( (p-p')^2 + (q-q')^2 + 2i(qp' - q'p) \right) \right)
\end{eqnarray}
We are now in a position to present equations (\ref{eq:thefinaleqnforkernsinctsof}) for the canonical transformation
\begin{equation} \nonumber
q \rightarrow -P \hspace{2cm} p \rightarrow Q.
\end{equation}
Using equations (\ref{eq:goodformofflipctone}), (\ref{eq:goodformofflipcttwo}), (\ref{eq:actiofcreatonctsone}) and  (\ref{eq:actiofanihilonctsone}) we can see that equations (\ref{eq:thefinaleqnforkernsinctsof}) must take the form
\begin{eqnarray} \nonumber
\lefteqn{\int_{\Space{R}{2n}} m(q,p,q'',p'') (q''+ip'')} \\ \nonumber
&& \hspace{0.7cm} \times \exp \left( -\frac{\pi}{2h} [(p'' -p')^2 + (q''-q')^2 + 2i(q''p'-q'p'')] \right) \, dq'' \, dp'' \\ \nonumber
&& = \int_{\Space{R}{2n}} m(q'',p'',q',p') i(q+ip) \\ \nonumber
&& \hspace{1cm} \times \exp \left( -\frac{\pi}{2h} [ (p-p'')^2 + (q-q'')^2 +2i(qp''-q''p) ] \right)  \, dq'' \, dp''.
\end{eqnarray}
and
\begin{eqnarray} \nonumber
\lefteqn{\int_{\Space{R}{2n}} m(q,p,q'',p'') (q'-ip')} \\ \nonumber
&& \hspace{0.5cm} \times \exp \left( - \frac{\pi}{2h} [(p'' -p')^2 + (q''-q')^2 + 2i(q''p'-q'p'')] \right) \, dq'' \, dp'' \\ \nonumber
&& = \int_{\Space{R}{2n}} m(q'',p'',q',p') (-i(q''-ip'')) \\ \nonumber
&& \hspace{1.2cm} \times \exp \left( -\frac{\pi}{2h} [ (p-p'')^2 + (q-q'')^2 +2i(qp''-q''p) ] \right)  \, dq'' \, dp''
\end{eqnarray}
for this canonical transformation. The function
\begin{eqnarray} \nonumber
m(q,p,q',p') = \exp \left( -\frac{\pi}{2h} ( q^2 +p^2 + q'^2 +p'^2 -2iqp -2iqq' +2iqp') \right)
\end{eqnarray}
can be shown to satisfy these equations through the repeated use of formulas (\ref{eq:waveletwith}) and (\ref{eq:waveletwithx}). Even though we have only looked at this equation for a linear example it must be stressed that it holds for non-linear examples also. We don't give any examples of this here as in the next section we derive some more manageable equations using the kernel states.

\subsection{Equations for non-linear transformations for states realised as kernels} \label{sect:nonlintransforstatekerns}

In \cite{Brodlie02} we showed that $p$-mechanical states could be realised as integration kernels. In this section we derive an equation similar to (\ref{eq:thefinaleqnforkernsinctsof}) for the kernel states. It is shown that this equation in many circumstances is easier to solve than (\ref{eq:thefinaleqnforkernsinctsof}).

Let $\kct$ denote the operator on the algebra of $p$-mechanical observables corresponding to a canonical transformation
\begin{equation} \nonumber
\kct B = U^{-1} B U.
\end{equation}
The adjoint operator $\kct^*$ action on a kernel $l$ is defined by
\begin{equation} \label{eq:ctandtheprimfuncobsvrel}
\langle \kct B, l \rangle = \langle B, \kct^* l \rangle.
\end{equation}
Note here that this is not an inner product, instead a functional on the right acting on a $p$-observable which is on the left. In section \ref{sect:introtopmech} we showed that any kernel $l$ can be expanded using the coherent state kernels, that is
\begin{equation} \nonumber
l(s,x,y) = \int_{\Space{R}{2n}} \int_{\Space{R}{2n}} l(s,x',y') \overline{l_{(h,q,p)} (s,x',y')} \, dx' \, dy' l_{(h,q,p)} (s,x,y) dq dp.
\end{equation}
We now derive an integral equation which when solved will give us
$\langle \kct^* l_{(h,q,p)} , l_{(h,q',p')} \rangle$. Initially we present a Lemma which will give an exact formula for the $p$-mechanisation of a classical observable evaluated by a kernel coherent state.

\begin{Lemma} \label{lem:aclassobsandakercs}
If $f$ is a classical observable and $\proj$ is the map of $p$-mechanisation as defined in equation (\ref{eq:pmechmap}) then
\begin{eqnarray} \nonumber
\lefteqn{\langle \proj (f) , \lqp \rangle} \\ \nonumber
&& = \left( \frac{2}{h} \right)^n \exp \left( -\frac{2\pi}{h} (q^2 +p^2) \right) \\ \nonumber
&& \hspace{2cm} \times \int_{\Space{R}{2n}} f(a,b) \exp \left( -\frac{2\pi}{h} (a^2+b^2) + \frac{4\pi}{h} (aq+bp) \right) \, da \, db.
\end{eqnarray}
\end{Lemma}

\begin{proof}
By a direct calculation
\begin{eqnarray} \nonumber
\lefteqn{\langle \proj (f) , \lqp \rangle} \\ \nonumber
&& = \int_{\Space{R}{2n+1}} \delta(s) \, \int_{\Space{R}{2n}} f(a,b) \exp (2\pi i (ax+by)) \, da \, db \, \\ \nonumber
&& \qquad \qquad \times \exp \left( 2\pi ihs - 2\pi i (qx+py) -\frac{\pi h}{2}(x^2 +y^2) \right) \, ds \, dx \, dy
\end{eqnarray}
which implies that
\begin{eqnarray} \nonumber
\lefteqn{\langle \proj (f) , \lqp \rangle} \\ \nonumber
&& = \int_{\Space{R}{}} \delta(s) \exp (2 \pi ihs) \, ds \\ \nonumber
&& \hspace{1.5cm} \times \int_{\Space{R}{4n}} f(a,b) \exp (x(2\pi i (a-q))) \exp (y(2\pi i (b-p))) \\ \nonumber
&&  \hspace{4cm} \times \exp \left( -\frac{\pi h}{2} (x^2 + y^2 ) \right) \, da \, db \, dx \, dy.
\end{eqnarray}
Using (\ref{eq:waveletwithx}) the right hand side of the above equation becomes
\begin{eqnarray} \nonumber
\lefteqn{ \left( \frac{2}{h} \right)^n \int_{\Space{R}{2n}} f(a,b) \exp \left( \frac{ (\pi i (a-q))^2}{\pi h/2} \right) \exp \left( \frac{(\pi i (b-p))^2}{\pi h /2} \right) \, da \, db} \\ \nonumber
&& = \left( \frac{2}{h} \right)^n \int_{\Space{R}{2n}} f(a,b) \exp \left( -\frac{2\pi}{h} [ (a-q)^2 + (b-p)^2 ] \right) \, da \, db
\end{eqnarray}
The result follows from a trivial rearrangement of the above equation.
\end{proof}
So now if we have $2n$ relations as in (\ref{eq:newinitialconrelq}) and (\ref{eq:newinitialconrelp}) we can define the operator $\kct$ by the relation
\begin{equation} \nonumber
\kct \proj (F_i) = \proj (f_i) .
\end{equation}
Applying the kernel $\lqp$ to both sides of this equation we get
\begin{equation} \nonumber
\langle \kct \proj (F_i), \lqp \rangle = \langle \proj (f_i), \lqp \rangle.
\end{equation}
This is equivalent to
\begin{equation} \label{eq:ctforkernone}
\langle \proj (F_i) , \kct^* \lqp \rangle = \langle \proj (f_i) , \lqp \rangle.
\end{equation}
However
\begin{equation} \label{eq:ctforkerntwo}
\kct^* \lqp = \int_{\Space{R}{2n}} \langle \kct^* \lqp , \lqpp \rangle \lqpp \, dq' \, dp'.
\end{equation}
where the $\langle,\rangle$ for two kernels is just $\langle l,l' \rangle = \int_{\Space{R}{2n}} l \bar{l'} \, dx \, dy$ (however if they contain an observable and a kernel it is still the evaluation of the observable by the functional). Substituting (\ref{eq:ctforkerntwo}) into (\ref{eq:ctforkernone}) gives us
\begin{equation} \nonumber
\left\langle \proj (F_i) ,  \int_{\Space{R}{2n}} \langle \kct^* \lqp , \lqpp \rangle \lqpp \, dq' \, dp' \right\rangle = \langle \proj (f_i) , \lqp \rangle ,
\end{equation}
which is equivalent to
\begin{equation} \nonumber
\int_{\Space{R}{2n}} \langle \kct^* \lqp , \lqpp \rangle \langle \proj (F_i) , \lqpp \rangle \, dq' \, dp' = \langle \proj (f_i), \lqp \rangle.
\end{equation}
Using Lemma \ref{lem:aclassobsandakercs} this equation becomes
\begin{eqnarray} \nonumber
\lefteqn{\int_{\Space{R}{2n}} \langle \kct^* \lqp, \lqpd \rangle  \left( \frac{2}{h} \right)^n \exp \left( -\frac{2\pi}{h} (q'^2 +p'^2) \right)} \\ \nonumber
&& \hspace{1cm}  \times \int_{\Space{R}{2n}} F_i(a,b) \exp \left( -\frac{2\pi}{h} (a^2+b^2) + \frac{4\pi}{h} (aq'+bp') \right) \, da \, db \, dq' \, dp' \\ \nonumber
&& =  \left( \frac{2}{h} \right)^n \exp \left( -\frac{2\pi}{h} (q^2 +p^2) \right) \\ \nonumber
&& \hspace{1.2cm}  \times \int_{\Space{R}{2n}} f_i(a,b) \exp \left( -\frac{2\pi}{h} (a^2+b^2) + \frac{4\pi}{h} (aq+bp) \right) \, da \, db ,
\end{eqnarray}
which can be simplified to
\begin{eqnarray}  \label{eq:thebigctforkerns}
\lefteqn{\int_{\Space{R}{2n}} \langle \kct^* \lqp, \lqpd \rangle \exp \left( -\frac{2\pi}{h} (q'^2 +p'^2) \right)} \\ \nonumber
&& \hspace{0.3cm} \times \int_{\Space{R}{2n}} F_i(a,b) \exp \left( -\frac{2\pi}{h} (a^2+b^2) + \frac{4\pi}{h} (aq'+bp') \right) \, da \, db \, dq' \, dp' \\ \nonumber
&& = \exp \left( -\frac{2\pi}{h} (q^2 +p^2) \right) \\ \nonumber
&& \hspace{1.7cm} \times \int_{\Space{R}{2n}} f_i(a,b) \exp \left( -\frac{2\pi}{h} (a^2+b^2) + \frac{4\pi}{h} (aq+bp) \right) \, da \, db .
\end{eqnarray}
We will now go on to show that for a number of canonical transformations this integral equation takes a clear form which is easy to solve.

\subsection{The Hamilton Transformation from the Forced Oscillator}

We now demonstrate how equations (\ref{eq:thebigctforkerns}) can deal with a non-linear transformation. We do this through applying it to the Hamilton transformation for the forced oscillator. This is the canonical transformation which is generated by the time evolution of phase space due to the forced oscillator. The $p$-mechanical forced oscillator is discussed in \cite{Brodlie02}, for simplicity we consider the oscillator to be of unit mass and unit frequency, but forced by an arbitary function $z(t)$. The classical canonical transformation (this is for the time evolution from time $0$ to time $t$) is defined by
\begin{eqnarray} \nonumber
Q &=& q\cos(t) + p \sin (t) + \int_0^t z(\tau) \sin (\tau) \, d\tau \\ \nonumber
P &=& -q \sin (t) + p \cos (t) +\int_0^t z (\tau) \cos (\tau) \, d\tau.
\end{eqnarray}
Using equations (\ref{eq:waveletwith}) and  (\ref{eq:waveletwithx}) we get the relations
\begin{eqnarray} \nonumber
\exp \left( -\frac{2\pi}{h} (q^2 +p^2) \right) \int_{\Space{R}{2n}} a \exp \left( -\frac{2\pi}{h} (a^2+b^2) + \frac{4\pi}{h} (aq+bp) \right) \, da \, db &=& \frac{h}{2} q \\ \nonumber
\exp \left( -\frac{2\pi}{h} (q^2 +p^2) \right) \int_{\Space{R}{2n}} b \exp \left( -\frac{2\pi}{h} (a^2+b^2) + \frac{4\pi}{h} (aq+bp) \right) \, da \, db &=& \frac{h}{2} p \\ \nonumber
\exp \left( -\frac{2\pi}{h} (q^2 +p^2) \right) \int_{\Space{R}{2n}} \exp \left( -\frac{2\pi}{h} (a^2+b^2) + \frac{4\pi}{h} (aq+bp) \right) \, da \, db &=& \frac{h}{2}.
\end{eqnarray}
These relations imply that equations (\ref{eq:thebigctforkerns}) for this transformation take the form
\begin{eqnarray} \nonumber
\lefteqn{ \int_{\Space{R}{2n}} \langle \kct^* \lqp , \lqpp \rangle q' \, dq' \, dp'} \\ \label{eq:mmeqnforforcedoscone}
&& \qquad \qquad \qquad \qquad = q\cos(t) + p \sin (t) + \int_0^t z(\tau) \sin (\tau) \, d\tau \\ \nonumber
&& \int_{\Space{R}{2n}} \langle \kct^* \lqp , \lqpp \rangle p' \, dq' \, dp' \\ \label{eq:mmeqnforforcedosctwo}
&& \qquad \qquad \qquad \qquad = -q \sin (t) + p \cos (t) +\int_0^t z (\tau) \cos (\tau) \, d\tau.
\end{eqnarray}
By observing equations (\ref{eq:waveletwith}) and (\ref{eq:waveletwithx}), a potential solution of equations (\ref{eq:mmeqnforforcedoscone}) and (\ref{eq:mmeqnforforcedosctwo}) is
\begin{eqnarray} \label{eq:solnofkctforfosc}
\lefteqn{\langle \kct^* \lqp , \lqpp \rangle} \\ \nonumber
&=& \frac{1}{h} \exp \left( -\frac{\pi}{h} \left[ \left( q \cos (t) + p \sin (t) + \int_0^t z(\tau) \sin (\tau) \, d\tau -q' \right)^2 \right. \right. \\ \nonumber
&& \qquad \qquad \qquad \left. \left. + \left( \int_0^t z (\tau) \cos (\tau) \, d\tau - q \sin (t) + p \cos (t) - p' \right)^2 \right] \right).
\end{eqnarray}
We now show that this satisfies (\ref{eq:mmeqnforforcedoscone})
\begin{eqnarray}
\lefteqn{\int_{\Space{R}{2n}} \langle \kct^* \lqp , \lqpp \rangle q' \, dq' \, dp'} \\ \nonumber
&=& \int_{\Space{R}{2n}}  \frac{1}{h} \exp \left( -\frac{\pi}{h} \left[ \left( q \cos (t) + p \sin (t) + \int_0^t z(\tau) \sin (\tau) \, d\tau -q' \right)^2 \right. \right. \\ \nonumber
&& \qquad \qquad \qquad  \left. \left. + \left( \int_0^t z (\tau) \cos (\tau) \, d\tau - q \sin (t) + p \cos (t) - p' \right)^2 \right] \right) q' \, dq' \, dp' \\ \nonumber
&=& \frac{1}{h} \exp \left( -\frac{\pi}{h} \left[ \left(q \cos (t) + p \sin (t) + \int_0^t z(\tau) \sin (\tau) \, d\tau \right)^2 \right. \right. \\ \nonumber
&& \qquad \qquad \qquad \qquad \left. \left. + \left( \int_0^t z (\tau) \cos (\tau) \, d\tau - q \sin (t) + p \cos (t) \right)^2 \right] \right) \\ \nonumber
&& \times \int_{\Space{R}{2n}} \exp \left( \frac{2\pi}{h} \left[ q' \left( q \cos (t) + p \sin (t) + \int_0^t z(\tau) \sin (\tau) \, d\tau \right) \right. \right. \\ \nonumber
&& \qquad \qquad \qquad \qquad \left. \left. + p'\left( \int_0^t z (\tau) \cos (\tau) \, d\tau - q \sin (t) + p \cos (t) \right) \right] \right) \\ \nonumber
&& \qquad \qquad \qquad \qquad \qquad \qquad \qquad \qquad \times \exp \left( -\frac{\pi}{h} (q'^2+p'^2) \right) q' \, dq' \, dp'
\end{eqnarray}
Using (\ref{eq:waveletwith}) and (\ref{eq:waveletwithx}) this becomes
\begin{eqnarray} \nonumber
\lefteqn{\int_{\Space{R}{2n}} \langle \kct^* \lqp , \lqpp \rangle q' \, dq' \, dp'} \\ \nonumber
&=& \frac{1}{h} \exp \left( -\frac{\pi}{h} \left[ \left(q \cos (t) + p \sin (t) + \int_0^t z(\tau) \sin (\tau) \, d\tau \right)^2 \right. \right. \\ \nonumber
&& \qquad \qquad \qquad \qquad \left. \left. + \left( \int_0^t z (\tau) \cos (\tau) \, d\tau - q \sin (t) + p \cos (t) \right)^2 \right] \right) \\ \nonumber
&& \times h \left(q \cos (t) + p \sin (t) + \int_0^t z(\tau) \sin (\tau) \, d\tau \right) \\ \nonumber
&& \qquad \qquad \times \exp \left( \frac{\pi}{h} \left[ \left( q \cos (t) + p \sin (t) + \int_0^t z(\tau) \sin (\tau) \, d\tau \right)^2 \right. \right. \\ \nonumber
&& \left. \left. \qquad \qquad \qquad \qquad + \left( \int_0^t z (\tau) \cos (\tau) \, d\tau - q \sin (t) + p \cos (t) \right)^2 \right] \right) \\ \nonumber
&=& q \cos (t) + p \sin (t) + \int_0^t z(\tau) \sin (\tau) \, d\tau .
\end{eqnarray}
By a similar calculation we can show that (\ref{eq:solnofkctforfosc}) satisfies (\ref{eq:mmeqnforforcedoscone}).

\subsection{A note on non-bijective transformations} \label{sect:nonbijcts}

In \cite{MoshinskySeligman78} the problem of representing non-bijective canonical transformations in quantum mechanics is considered. The majority of canonical transformations in classical mechanics are non-bijective --- one example is the action angle variables for the Kepler problem.  If we can represent non-bijective canonical transformations in $p$-mechanics we can use the infinite dimensional representations to get their representation in quantum mechanics. The physical importance of non-bijective canonical transformations in quantum mechanics is discussed in \cite[Sect. 7]{MoshinskySeligman78}. It is claimed that some non-linear canonical transformations can be used to show that some elements of quantum mechanics are already contained in classical mechanics.

We now outline a method of how to deal with non-bijective canonical transformations in $p$-mechanics. Our method is best illustrated through an example --- we look at the transformation into the action angle co-ordinates for the repulsive oscillator. This is a non-linear, non-bijective transformation which is discussed in great detail in \cite{MoshinskySeligman78}. The canonical transformation is
\begin{eqnarray} \label{eq:reposccttwo}
Q &=& \ln | p+q | , \\ \label{eq:reposcctone}
P &=& \frac{1}{2} ( p^2 - q^2 ) .
\end{eqnarray}
The non-bijectiveness of this transformation is manifested by the points $(q,p)$ and $(-q,-p)$ in the original phase space being mapped into the same point. Also the entire line $q+p=0$ is mapped to the single point $Q=-\infty,P=0$. To derive an equation for the states realised as kernels we put equations (\ref{eq:reposccttwo}) and  (\ref{eq:reposcctone}) into the form
\begin{eqnarray} \nonumber
\exp(2 Q) &=& (p+q)^2 , \\ \nonumber
P &=& \frac{1}{2} (p^2 - q^2).
\end{eqnarray}
To derive equations (\ref{eq:thebigctforkerns}) for this example we need the following Lemma
\begin{Lemma} \label{lem:pmechforreposccts}
We have the following relations
\begin{eqnarray} \nonumber
&& \exp \left( -\frac{2\pi}{h} (q^2 +p^2) \right) \int_{\Space{R}{2n}} \exp(2a) \exp \left( -\frac{2\pi}{h} (a^2+b^2) + \frac{4\pi}{h} (aq+bp) \right) \, da \, db  \\  \label{eq:expincsandkern}
&& \hspace{1cm} = \left( \frac{h}{2} \right) \exp \left( 2q + \frac{h}{2\pi} , \right)
\end{eqnarray}
\begin{eqnarray} \nonumber
&& \exp \left( -\frac{2\pi}{h} (q^2 +p^2) \right) \int_{\Space{R}{2n}} a^2 \exp \left( -\frac{2\pi}{h} (a^2+b^2) + \frac{4\pi}{h} (aq+bp) \right) \, da \, db  \\ \label{eq:squaredincsandkern}
&& \hspace{1cm} = \frac{h}{2} \left(\frac{h}{4\pi} + q^2  \right).
\end{eqnarray}
Clearly analogous relations to these hold if we replace $a$ by $b$ on the left hand side and $q$ by $p$ on the right hand side. Furthermore we have that
\begin{eqnarray} \nonumber
&& \exp \left( -\frac{2\pi}{h} (q^2 +p^2) \right) \int_{\Space{R}{2n}} ab \exp \left( -\frac{2\pi}{h} (a^2+b^2) + \frac{4\pi}{h} (aq+bp) \right) \, da \, db  \\ \label{eq:qpincsandkern}
&& \hspace{1cm} = \frac{h}{2} qp.
\end{eqnarray}
\end{Lemma}
\begin{proof}
Equation (\ref{eq:expincsandkern}) follows from a direct calculation using (\ref{eq:waveletwithx})
\begin{eqnarray} \nonumber
\lefteqn{ \exp \left( -\frac{2\pi}{h} (q^2 +p^2) \right) \int_{\Space{R}{2n}} \exp (2a) \exp \left( -\frac{2\pi}{h} (a^2+b^2) + \frac{4\pi}{h} (aq+bp) \right) \, da \, db} \\ \nonumber
&& = \exp \left( -\frac{2\pi}{h} (q^2 +p^2) \right) \left( \frac{h}{2} \right) \exp \left( \frac {2 \pi p^2}{h} \right) \exp \left( \frac{ \left( \frac{2\pi}{h} q +1 \right)^2}{2\pi / h} \right) \\ \nonumber
&& = \left(\frac{h}{2}\right) \exp \left( 2q + \frac{h}{2\pi} \right)
\end{eqnarray}
Similarly (\ref{eq:squaredincsandkern}) can be verified by a direct calculation using (\ref{eq:waveletwithxforalln}) with $n=2$. Likewise (\ref{eq:qpincsandkern}) can be verified by using (\ref{eq:waveletwith}) twice.
\end{proof}

Using Lemma \ref{lem:pmechforreposccts} equations (\ref{eq:thebigctforkerns}) for this example take the form
\begin{eqnarray} \label{eq:cteqnforkernsforreposcone}
\int_{\Space{R}{2n}} \langle \kct^* \lqp , \lqpp \rangle \exp \left( 2q' + \frac{h}{2\pi} \right) \, dq' \, dp' = (p+q)^2 + \frac{h}{2\pi} \\ \label{eq:cteqnforkernsforreposctwo}
\int_{\Space{R}{2n}} \langle \kct^* \lqp , \lqpp \rangle  p' \, dq' \, dp' = (p^2 - q^2).
\end{eqnarray}
The non-bijectiveness is apparent in equations (\ref{eq:cteqnforkernsforreposcone}) and (\ref{eq:cteqnforkernsforreposctwo}) since they are invariant under the translation $(q,p) \mapsto (-q,-p)$.
Any solution of (\ref{eq:cteqnforkernsforreposcone}) and (\ref{eq:cteqnforkernsforreposctwo}) will be such that
\begin{equation} \label{eq:evenforreposc}
\langle \kct^* \lqp , \lqpp \rangle = \langle \kct^* l_{(h,-q,-p)} , \lqpp \rangle.
\end{equation}
Since
\begin{equation} \nonumber
\kct^* \lqp = \int_{\Space{R}{2n}} \langle \kct^* \lqp , \lqpp \rangle \lqpp \, dq' \, dp'
\end{equation}
it is clear that $\kct^* \lqp = \kct^* l_{(h,-q,-p)}$. Now we show that this map $\kct^* : \sokh \rightarrow \sokh$ is not a bijection.
We have the formula for any kernel $l$
\begin{equation} \nonumber
\kct^* l =  \int_{\Space{R}{2n}} \int_{\Space{R}{2n}} \langle \kct^* \lqp, \lqpp \rangle \langle l, \lqp \rangle \lqpp \, dq \, dp \, dq' \, dp'.
\end{equation}
So $\kct^* l = \kct^* l'$ if and only if
\begin{equation} \nonumber
\int_{\Space{R}{2n}} \langle \kct^* \lqp, \lqpp \rangle \langle l, \lqp \rangle \, dq \, dp = \int_{\Space{R}{2n}} \langle \kct^* \lqp, \lqpp \rangle \langle l', \lqp \rangle \, dq \, dp
\end{equation}
holds for almost all\footnote{Almost every and almost everywhere are measure theoretic terms, see \cite{ReedSimon80}, \cite{KolmogorovFomin75}.} $q',p'$. By (\ref{eq:evenforreposc}) this is equivalent to
\begin{eqnarray} \nonumber
&& \int_{q+p>0} (\langle l, \lqp \rangle + \langle l , \lqpm \rangle) \langle \kct^* \lqp, \lqpp \rangle \, dq \, dp \\  \nonumber
&& \qquad \qquad = \int_{q+p>0} (\langle l', \lqp \rangle + \langle l' , \lqpm \rangle) \langle \kct^* \lqp, \lqpp \rangle \, dq \, dp.
\end{eqnarray}
This holds if and only if
\begin{equation} \label{eq:equivrelfornonbij}
\langle l, \lqp \rangle + \langle l , \lqpm \rangle = \langle l', \lqp \rangle + \langle l' , \lqpm \rangle
\end{equation}
holds for almost every $q,p$ such that $q+p > 0$. So one way of restoring bijectiveness would be to reduce the size of the original set of kernels by factoring out by the equivalence relation, $l \sim l'$ if and only if (\ref{eq:equivrelfornonbij}) holds. In the obvious way we would have a bijection from $\sokh / \sim$ to $\sokh$.

Instead of taking this approach we take the simpler approach of increasing the size of the set of kernels to which we map. In doing so we derive a bijection from $\sokh$ to $\sokh \times \sokh$. We do this through introducing two new operators $\kctsp, \kctsm: \sokh \rightarrow \sokh$ which are defined as
\begin{displaymath}
\kctsp \lqp = \left\{
\begin{array}{lll}
\kct^* \lqp  & \textrm{ , if  } & q+p>0; \\
0  & \textrm{ , if  } & q+p<0;
\end{array} \right.
\end{displaymath}
and
\begin{displaymath}
\kctsm \lqp = \left\{
\begin{array}{lll}
0  & \textrm{ , if  } & q+p>0; \\
\kct^* \lqp  & \textrm{ , if  } & q+p<0 .
\end{array} \right.
\end{displaymath}
These operators can be extended to the entire space by coherent state (i.e. wavelet) expansions. Furthermore we define the mapping $: \kctst : \sokh \rightarrow \sokh \times \sokh$ by
\begin{equation} \nonumber
\kctst l = [ \kctsp l, \kctsm l ].
\end{equation}
So for any kernel $l$
\begin{eqnarray} \nonumber
\kctst l &=& \int_{\Space{R}{2n}} \langle l, \lqp \rangle \kctst \lqp \, dq \, dp \\ \nonumber
&=& \int_{\Space{R}{2n}} \langle l, \lqp \rangle [ \kctsp \lqp , \kctsm \lqp ] \, dq \, dp \\ \nonumber
&=& [ \int_{q+p>0} \langle l, \lqp \rangle \kct^* \lqp \, dq \, dp ,  \int_{q+p<0} \langle l, \lqp \rangle \kct^* \lqp \, dq \, dp ].
\end{eqnarray}
Hence $\kctst l = \kctst l'$ if and only if
\begin{equation} \nonumber
\int_{q+p>0} \langle l, \lqp \rangle \kct^* \lqp \, dq \, dp = \int_{q+p>0} \langle l', \lqp \rangle \kct^* \lqp \, dq \, dp
\end{equation}
and
\begin{equation} \nonumber
\int_{q+p<0} \langle l, \lqp \rangle \kct^* \lqp \, dq \, dp = \int_{q+p<0} \langle l', \lqp \rangle \kct^* \lqp \, dq \, dp.
\end{equation}
These two equations hold if and only if
\begin{equation} \nonumber
\langle l, \lqp \rangle = \langle l', \lqp \rangle
\end{equation}
holds for almost every $q,p$ such that $q+p>0$ and almost every $q,p$ such that $q+p<0$. This is equivalent to $l$ and $l'$ being equal. Hence we have restored bijectiveness.

The implications to quantum mechanics of this are as follows. We now have a map $\kctt$ (the adjoint of $\kctst$) which transforms $p$-mechanical observables corresponding to this non-bijective canonical transformation. If we take the $\rho_h$ representation of this we get a map for quantum observables.

This work also has classical implications. Initially looking at non-statistical mechanics we have if the canonical transformation changes a classical observable $f(q,p)$ into $\tilde{f} (q,p)$ and the $p$-mechanisation of $f$ is $B(s,x,y)$ then
\begin{eqnarray} \nonumber
\tilde{f} (q,p) = \langle B , \kct^* \lqpo \rangle = \langle B, \kct^* \lqpom \rangle = \tilde{f} (-q,-p).
\end{eqnarray}
Hence we have a demonstration of the classical non-bijectiveness directly from $p$-mechanics. If we let $\mathrm{A}$ denote the set of classical observables, we have shown that the mapping from $\mathrm{A} \rightarrow \mathrm{A}$ by $f \mapsto \tilde{f}$ is non-bijective. However we can restore bijectiveness in the classical mapping through $p$-mechanics. If we introduce the operator $\kctstcl:\mathrm{A} \rightarrow \mathrm{A} \times \mathrm{A}$ by
\begin{equation} \nonumber
\kctstcl f (q,p)= \langle B, \kctst \lqpo \rangle
\end{equation}
where $B$ is the $p$-mechanisation of $f$, then
\begin{eqnarray} \nonumber
\kctstcl f(q,p) &=& \langle B, \kctst \lqpo \rangle \\ \nonumber
&=& [ \langle B, \kctsp \lqpo \rangle , \langle B, \kctsm \lqpo \rangle ] \\ \nonumber
&=& \left\{ \begin{array}{lll}
[\tilde{f} (q,p),0] & \textrm{  if  } & q+p>0 \\ \\[0.0mm]
[ 0 , \tilde{f} (q,p) ] & \textrm{  if  } & q+p<0 .
\end{array} \right.
\end{eqnarray}
Now we have a map $\mathrm{A} \rightarrow \mathrm{A} \times \mathrm{A}$ representing this canonical transformation which is bijective. We can extend all of this to statistical mechanics \cite{Honerkamp98} by using linear combinations of these coherent states.

\section{Summary and Possible Extensions} \label{sect:summofctpaper}

     One of the main features of this work is demonstrating how using  coherent states in a Segal-Bargmann-Fock type space can sometimes be advantageous over using the matrix elements of position and momentum in $\ltworn$. Another feature of our equations is that they do not rely on the property that observables are elements of the algebra generated by the position and momentum operators. In \cite{MoshinskySeligman78, MelloMoshinsky75} all the quantum mechanical operators are derived using this algebra condition --- in this paper we use an integral transform instead. This integral transform at first makes our equations look less desirable but it is shown that for many examples they take a simple form. This work has also demonstrated the advantages of representing states as integration kernels --- this complements the work in \cite{Brodlie02}.

The most immediate extension of this work would be to look at more complex examples especially some more non-linear, non-bijective examples. One possible and interesting extension would be to extend these ideas to phase spaces other than $\Space{R}{2n}$. This would be to extend these ideas to a phase space which is a general symplectic manifold \cite[Chap. 5]{MarsdenRatiu99}, for example $T^*M$ for some general manifold $M$ \cite[Chaps. 7-10]{Arnold90}, \cite[Chap. 5]{Jose98}. Another interesting extension would be to look at the role of Egorov's theorem \cite{Kisil02.2} in infinitesimal canonical transformations for $p$-mechanics. Egorov's theorem \cite{Egorov86} has always been posed in the language of pseudodifferential and Fourier integral operators on $\ltworn$ this idea could be extended to our space $\fock$ with pseudodifferential operators being replaced by Toeplitz operators as in \cite{Howe80.1}.

All these equations could be defined outside the world of $p$-mechanics. If the usual coherent states in $\ltworn$ \cite[Sect. 10.7]{Merzbacher70} were used instead of the eigenfunctions of position and momentum Moshinsky's equations would be immediately tansformed into integral equations.

\section*{Acknowledgements}

I would like to thank my supervisor Dr V. Kisil for numerous discussions along with a great deal of encouragement. I would also like to thank the anonymous referee for some useful comments. The author is a PhD student supported by an EPSRC grant.

\bibliography{everything}

\bibliographystyle{plain}

\end{document}